\documentclass{ifacconf}

\usepackage{graphicx}      
\usepackage{natbib}        
\usepackage{bbm}
\usepackage{verbatim}
\usepackage{amsmath}
\usepackage{amsfonts}
\usepackage{bm}
\usepackage{float}

\usepackage{subcaption}

\let\theoremstyle\relax
\usepackage{amsthm}
\usepackage{mathrsfs}
\usepackage{xcolor}

\usepackage{todonotes}

\theoremstyle{definition}

\theoremstyle{plain}

\let\theoremstyle\relax
\usepackage{amsthm}

\newtheorem{definition}{Definition}

\newtheorem{remark}{Remark}

\newtheorem{problem}{Problem}

\DeclareMathOperator*{\minimize}{Minimize}
\DeclareMathOperator*{\maximize}{Maximize}

\begin{document}
\begin{frontmatter}

\title{A Mobility Equity Metric for Multi-Modal Intelligent Transportation Systems\thanksref{footnoteinfo}}

\thanks[footnoteinfo]{This research was supported by NSF under Grants CNS-2149520 and CMMI-2219761.}

\author{Heeseung Bang, Aditya Dave, Filippos N. Tzortzoglou,}
\author{and Andreas A. Malikopoulos} 

\address{School of Civil and Environmental Engineering, Cornell University, Ithaca, NY 14850, USA \\(emails: \{hb489, a.dave, ft253, amaliko\}@cornell.edu)}

\begin{abstract}                
In this paper, we introduce a metric to evaluate the equity in mobility and a routing framework to enhance the metric within multi-modal intelligent transportation systems.
The mobility equity metric (MEM) simultaneously accounts for service accessibility and transportation costs to quantify the equity and fairness in a transportation network.
Finally, we develop a system planner integrated with MEM that aims to distribute travel demand for the transportation network, resulting in a socially optimal mobility system.
Our framework results in a transportation network that is efficient in terms of travel time, improves accessibility, and ensures equity in transportation.

\end{abstract}


\end{frontmatter}

\section{Introduction}

With the rapid urbanization and population growth in modern cities, ensuring equitable access to services has become a significant concern. The ability to move using different modes of transportation is a fundamental necessity for all members of society to reach opportunities, amenities, and other necessities of life.
Consequently, there has been increasing research interest in the analysis of mobility equity (see 
\citet{guo2020systematic,litman2017evaluating} and reference therein).
This research has provided approaches to evaluate equity across different population demographics in terms of access to specific services, such as hospitals, jobs, etc., using a specific mode of transportation, usually traditional public transit.
However, with the emergence of intelligent transportation systems in modern cities, it is also important to study their impact on mobility equity. 

Intelligent transportation systems such as connected and automated vehicles (CAVs), on-demand mobility, or shared mobility systems can efficiently manage travel requests while alleviating traffic congestion, reducing energy consumption, and improving safety.
For instance, a series of research efforts have addressed the coordination problem of CAVs in various traffic scenarios to reduce energy consumption with guaranteed safety \citep{Au2015,li2018game,zhou2019development,bang2024confidence}. Similarly, some studies have explored efficient routing frameworks for on-demand mobility systems \citep{ammous2017optimal,salazar2019congestion,wollenstein2021routing,bang2021AEMoD} and ride-sharing \citep{ta2017efficient,tsao2019model,yu2021optimal}. However, despite these extensive research efforts, there is a lack of equity consideration.
This could lead to the benefits of CAVs being distributed unevenly among people from different demographics, potentially exacerbating existing inequities. 
Therefore, there is a critical need to investigate and maximize the positive impact of intelligent transportation systems on existing societal inequities.

In this paper, we propose an improved mobility equity metric (MEM) that accounts for multi-modal emerging mobility systems and accommodates multiple services simultaneously.
We first define \textit{mobility index} (MI) that measures accessibility to essential services while accounting for travel costs, and then capture how mobility indices are well distributed across a city by using a Gini coefficient \citep{gastwirth1972estimation}.
This MI only requires publicly available data, and we present a method of measuring MI using isochrone and point-of-interest (POI) datasets.
Then, we design a system planner to provide route suggestions to the vehicles in the network to improve MEM. The system accommodates private vehicles that may or may not comply with the planner's suggestions.
The main contributions of this paper are:
(1) the introduction of MEM to evaluate the distribution of the \textit{ability to move} across different regions; (2) the development of a method to improve MEM in a multi-modal intelligent transportation system; and (3) an analysis of the impact of the system-planner on various types of travelers through numerical simulations.

The remainder of this paper is organized as follows.
In Section \ref{section:mem_intro}, we introduce the MEM and explain a method to evaluate it using real data. In Section \ref{sec:control}, we formulate a routing problem and incorporate it with MEM optimization problem.
We conduct numerical simulations and analyze results in Section \ref{sec:simulation}, and finally, in Section \ref{sec:conclusion}, we discuss concluding remarks and possible future work.

\section{Mobility Equity Measurement} \label{section:mem_intro}

In this section, we introduce the notion of mobility equity and explain our approach towards quantifying it. When analyzing mobility within a transportation network, it is important to account for a variety of sociotechnical factors, e.g., the financial demographics, places of origin, potential destinations, and the purpose of a trip. Similarly, important factors pertaining to the network include the different modes of transportation available to any traveler and the costs and travel times associated with using different modes to reach any given destination. A good measure of mobility equity must simultaneously incorporate these diverse factors into one measure across the complete network.


To this end, we consider a transportation network denoted by a directed graph $\mathcal{G} = (\mathcal{V},\mathcal{E})$, where $\mathcal{V}\subset\mathbb{N}$ denotes a set of nodes and $\mathcal{E}\subset\mathcal{V}\times\mathcal{V}$ constitutes a set of edges. The set $\mathcal{M}$ collects various modes of transportation available to travelers, e.g., public transportation, shared mobility, private vehicles, cycling, and walking. The set $\mathcal{S}$ collects the different types of services accessible through the transportation network. For the purpose of measuring equity, we restrict attention to travel pertaining to access to different services within the network. Such service types include medical facilities, financial services, groceries, schools, entertainment venues, restaurants, etc. The selection of these service types can be either specialized or generalized to measure mobility equity in different situations, e.g., we could consider travel pertaining exclusively to employment opportunities as in \cite{deboosere2018evaluating}.

For each mode $m\in\mathcal{M}$, $c_m$ denotes the cost per passenger mile incurred by travelers. For each service type $s \in \mathcal{S}$, the priority level is given by $\beta^s$, where a higher priority level indicates a more important service. When quantifying mobility equity, we prioritize accessibility to essential services such as medical facilities, groceries, financial services, and schools over non-essential services such as entertainment venues and shopping malls. 
Furthermore, the time threshold $\tau_m$ represents a reasonable limit on the time a traveler would spend accessing different services using mode $m$. Then, for a node $i \in \mathcal{V}$ of the transportation network, $\sigma_{i,m}^s(\tau_m)$ represents the count of services of type $s \in \mathcal{S}$ accessible within time $\tau_m$ using mode $m$ starting at node $i$. Next, we show how these notions combine to provide a measure of \textit{accessibility}, known as the mobility index (MI), for travelers originating at any $i \in \mathcal{V}$.

\begin{definition} \label{def:mobility_index}
\vspace{8pt}
For a given network $\mathcal{G}$ with modes $\mathcal{M}$ and services $\mathcal{S}$, the mobility index (MI) at node $i\in\mathcal{V}$ is
\begin{equation} \label{eq:MI}
    \varepsilon_i = \sum_{m\in\mathcal{M}} e^{-\kappa_i c_m} \cdot \left\{ \sum_{s\in\mathcal{S}} \beta^s \sigma_{i,m}^s(\tau_m) \right\},
\end{equation}
where $\kappa_i$ is a constant of price sensitivity, capturing the trade-off between travel cost and travel time.
\end{definition}

The MI introduced in Definition \ref{def:mobility_index} incorporates travel time for different service types using the inner term and travel cost and price sensitivity using the outer, exponentially decaying term.
In the inner term, we sum up the number of accessible services after weighing them according to their priority levels to ensure that our metric accounts for various reasons of travel. 
In the outer term, the price sensitivity $\kappa_i$ varies with the choice of node $i \in \mathcal{V}$ to accommodate different preferences and socio-economic factors associated with travelers originating from different nodes within a transportation network.
The differences in the trade-off between travel time and travel cost for different types of travelers can be captured by appropriately selecting $\kappa_i$ for each $i \in \mathcal{V}$.
In a neighborhood $i$ with a high price sensitivity $\kappa_i$, the MI $\epsilon_i$ naturally rewards an increase in number of accessible services $\sum_{s \in \mathcal{S}}\beta^{s}\sigma_{i,m}^{s}(\tau_m)$ for a low-cost mode, e.g., public transit, \textit{more} than an increase in number of accessible services using a high-cost mode, e.g., private vehicles.
In doing so, MI $\varepsilon_i$ measures the \textit{ability to move} for a traveler originating from any node $i \in \mathcal{V}$. However, it is important to recognize that the choice of hyperparameter values $\big(\tau_m, \beta^s, \kappa_i: m \in \mathcal{M}, s \in \mathcal{S}, i \in \mathcal{V} \big)$ plays an important role and the selection of good hyperparameter values in \eqref{eq:MI} is a subject of ongoing research.

\begin{figure}
    \centering
    \includegraphics[width=0.8\linewidth]{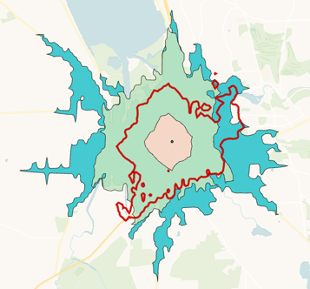}
    \caption{Isochrone of different transportation modes at Ithaca, NY, USA.}
    \label{fig:isochrone}
\end{figure}





\begin{figure*}
    \centering
    \subfloat[]{\includegraphics[height=4.1cm]{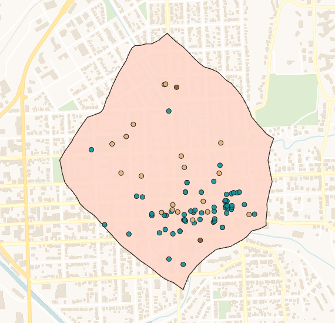}}
    \subfloat[]{\includegraphics[height=4.1cm]{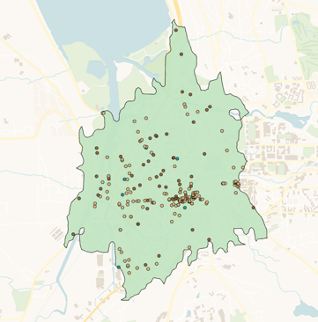}}
    \subfloat[]{\includegraphics[height=4.1cm]{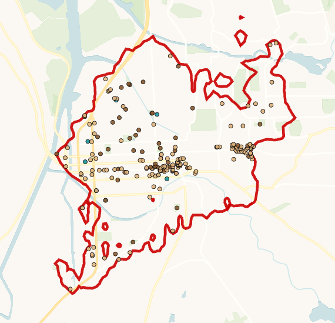}}
    \subfloat[]{\includegraphics[height=4.1cm]{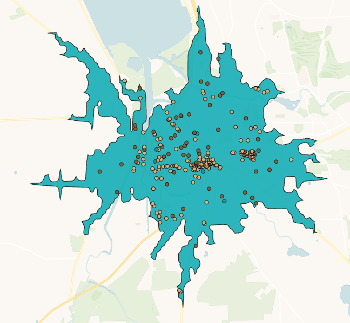}}
    \caption{Illustration of accessible services within the isochrone for each mode of transportation: (a) walking, (b) bicycle, (c) public transit, (d) driving.}
    \label{fig:services}
\end{figure*}

\subsection{Mobility Index Evaluation}

The evaluation of MI $\varepsilon_i$ for any node $i \in \mathcal{V}$ within a transportation network $\mathcal{G}$ begins with a computation of the number of accessible services $\sigma_{i,m}^{s}(\tau_m)$ of each type $s \in \mathcal{S}$ using each mode $m \in \mathcal{M}$. To achieve this, we leverage publicly available isochrone data. Each isochrone is a polygon surrounding node $i$ depicting the geographical boundary of the region accessible from $i$ using mode $m$ within a time threshold $\tau_m$.
Figure \ref{fig:isochrone} illustrates isochrone maps generated for various transportation modes (private car, public transit, bicycle, and walking) while considering approximate traffic in Ithaca, New York. For these isochrones, we select a time threshold of $20$ minutes for public transit and a threshold of $10$ minutes for cars, bicycles, and walking.

To collect the numbers of services of various types within different regions, we utilize a publicly available POI dataset \cite{TOMTOM}. Each POI datum contains the geographic location of a specific service and its service type. By superimposing this data onto the isochrones, we can count the number of accessible services within specific time frames. 
Figure \ref{fig:services} visualizes the distribution of essential and non-essential services within an isochrone for each mode of transportation. The points represent different service locations, with color variations indicating the types of services. Essential services, such as medical facilities and schools, are marked in distinct colors.

\subsection{Equity Evaluation}

While the MI offers an approach to quantify the accessibility and cost of mobility from any node of a transportation network, evaluating equity across the network requires us to compare among the different nodes.
For example, analyzing an urban environment would imply comparing MI of different neighborhoods.
To this end, we employ a Gini coefficient \citep{gastwirth1972estimation} modified to include different populations at different origins in a transportation network.

\begin{definition} \label{def:mem}
    \vspace{3pt}
    Given MI $\varepsilon_i$ for each node $i\in\mathcal{V}$ of transportation network $\mathcal{G}$, the mobility equity metric (MEM) is 
    \begin{equation} \label{eq:mem}
        \mathrm{MEM}(\mathcal{G}) = 1-\frac{\sum_{i\in\mathcal{V}} \sum_{j\in\mathcal{V}} p_i p_j \left|\varepsilon_i - \varepsilon_j\right|}{2\big(\sum_{i\in\mathcal{V}} p_i\big)\big(\sum_{i\in\mathcal{V}} p_i \varepsilon_i\big)},
    \end{equation}
    where $p_i$ is the population of travelers at node $i \in \mathcal{V}$. 
\end{definition}

The MEM in \eqref{eq:mem} takes values in $[0,1]$, where $MEM(\mathcal{G}) = 1$ indicates the ideal case where MI $\varepsilon_i$ is exactly the same for all $i \in \mathcal{{V}}$, and $MEM(\mathcal{G}) = 0$ indicates the negative extreme of complete inequity.
This behavior of the MEM is a consequence of using the Gini coefficient to evaluate pairwise differences between MIs across all nodes in the network, which is $0$ when $\varepsilon_i = \varepsilon_j$ for all $i, j \in \mathcal{V}$. 
Note that when computing MEM, we weigh the MI for each node in the network with the population at the node. This better accounts for the discrepancies in MI and ensures that equity is measured with respect to the travelers within the network rather than simply the geographical location of the nodes. 

\begin{remark}
\vspace{3pt}
    In our prior work \citep{Bang2023mem}, MEM used the average MI across regions. However, the revised MEM provided in Definition \ref{def:mem} is superior because (a) it takes values in $[0,1]$, allowing comparisons between different networks, and (b) it is a true measure of equity that increases only if areas with low MI are improved.
\end{remark}

\begin{remark}
\vspace{3pt}
    In practice, when evaluating the MEM of a city's transportation network, we can only include finitely many nodes. However, the selection and omission of certain nodes can have a major impact on the eventual outcome of MEM evaluation. Thus, selecting a socially, economically, demographically, and geographically diverse set of nodes is important to represent the city's population.
\end{remark}

\begin{remark}
\vspace{3pt}
    When creating nodes for a city's transportation network, we consider one node per neighborhood. However, when creating nodes, it is crucial to explore different sizes (smaller or bigger) of spatial units for which the assumption of uniform price sensitivity for all individuals within that area remains tenable.
\end{remark}

\section{Control Framework} \label{sec:control}

In this section, we consider the problem of improving MEM using CAVs routed by a centralized system planner in a transportation network.
To achieve this goal, we utilize a routing framework that incorporates MEM optimization as an upper-level objective for a system planner and considers a time-optimal lower-level objective to determine individual routing suggestions. The structure of this framework is similar to \cite{Bang2023mem} but utilizes the MEM from \eqref{eq:mem} in the upper-level objective rather than relying upon simple averaging of MIs.

In this framework, we consider a simple transportation network with two modes of transportation: (1) Public transit comprising of CAVs, and (2) private vehicles that can be either automated or human-driven.
The public transit CAVs can be routed by a centralized system planner. However, the system planner can only recommend routes to private vehicles who may or may not comply with a given recommendation.
Thus, our goal is to offer system-wide MEM-optimizing routing for all public transit CAVs and compliant private vehicles while accounting for the impact generated in the network by non-compliant (privately owned) vehicles.

As in Section \ref{section:mem_intro}, the transportation network is represented by a directed graph $\mathcal{G} = (\mathcal{V},\mathcal{E})$, the set of modes is $\mathcal{M}$ and the set of service types is $\mathcal{S}$. Note that $\mathcal{M} = \{1,2\}$, where mode $1$ refers to public transit and mode $2$ refers to private vehicles. The set $\mathcal{N}=\{1,\dots,N\}$, $N\in\mathbb{N},$ represents a collection of possible trips, where each trip consists of an origin-destination pair. Each origin represents a candidate for travel demand within the network, and each destination represents a location that contains a certain number of services of various types.
Each trip $n \in \mathcal{N}$ is associated with the origin $o_n\in\mathcal{O} \subseteq \mathcal{V}$, destination $d_n\in\mathcal{D} \subseteq \mathcal{V}$. The corresponding \textit{compliant} travel demand rate for any trip $n \in \mathcal{N}$ using mode $m \in \mathcal{M}$ is $\alpha_{m,n}\in\mathbb{R}_{>0}$. This demand does not include the demands of non-compliant privately owned vehicles.

We formulate the routing problem by considering vehicle flows across the edges of the network. To this end, we define $x^{ij}_{m,n} \in \mathbb{R}_{\geq0}$ to be the flow of compliant vehicles on edge $(i,j) \in \mathcal{E}$ traveling from $o_n\in\mathcal{O}$ to $d_n\in\mathcal{D}$ with mode $m \in \mathcal{M}$.
The total complying-vehicle flow on edge $(i,j)$ is calculated by $x^{ij}=\sum_m\sum_n h_m \cdot x^{ij}_{m,n}$, where $h_m$ is the road occupancy of the mode $m$. For example, we consider occupancy of public transportation $h_\mathrm{public}=0.8$ whereas that of private vehicle $h_\mathrm{private}$ is $1$.
Then, we explicitly define the flow of non-compliant vehicles on edge $(i,j)$ by $q^{ij} \in \mathbb{R}_{\geq0}$. Recall that the non-compliant vehicles are comprised only of one mode, i.e., private vehicles.
Given the flows of both compliant and non-compliant vehicles on edge $(i,j)$, the travel time across the edge is estimated by the \textit{Bureau of Public Roads (BPR)} latency function:
\begin{equation}
    t^{ij}(x^{ij}+q^{ij}) = t^{ij}_0 \cdot \left( 1+0.15 \left(\frac{x^{ij}+q^{ij}}{\gamma^{ij}}\right)^4\right),
\end{equation}
where $t^{ij}_0 \in \mathbb{R}_{\geq0}$ is the free-flow travel time and $\gamma^{ij} \in \mathbb{R}_{\geq0}$ is capacity of the road on edge $(i,j)$.

\begin{figure*}[h!]
    \centering
    \begin{subfigure}[b]{0.31\textwidth}
        \includegraphics[width=\textwidth]{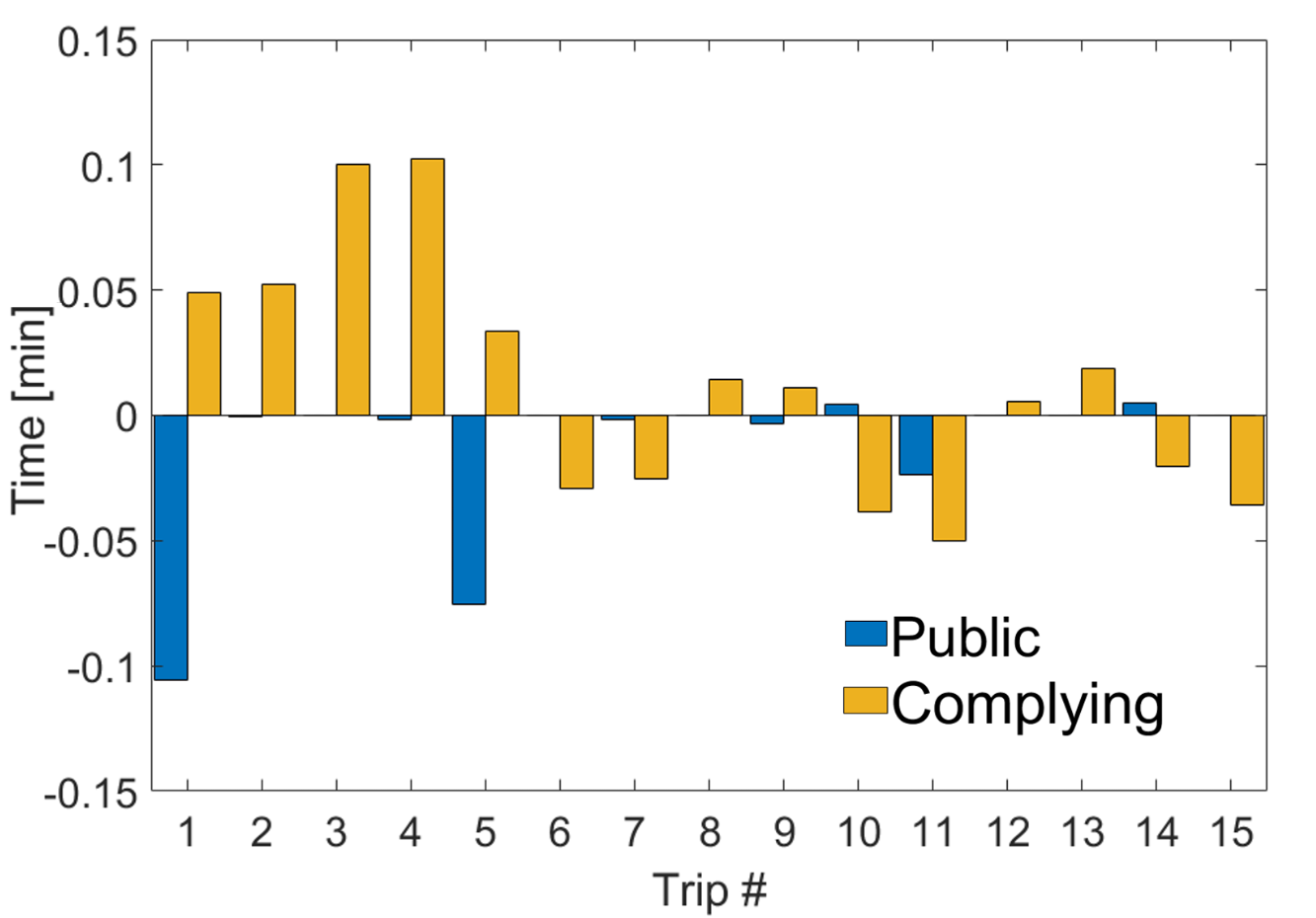}
             \caption{$10\%$ of non-compliance rate}
    \end{subfigure}
    \hfill
    \begin{subfigure}[b]{0.31\textwidth}
        \includegraphics[width=\textwidth]{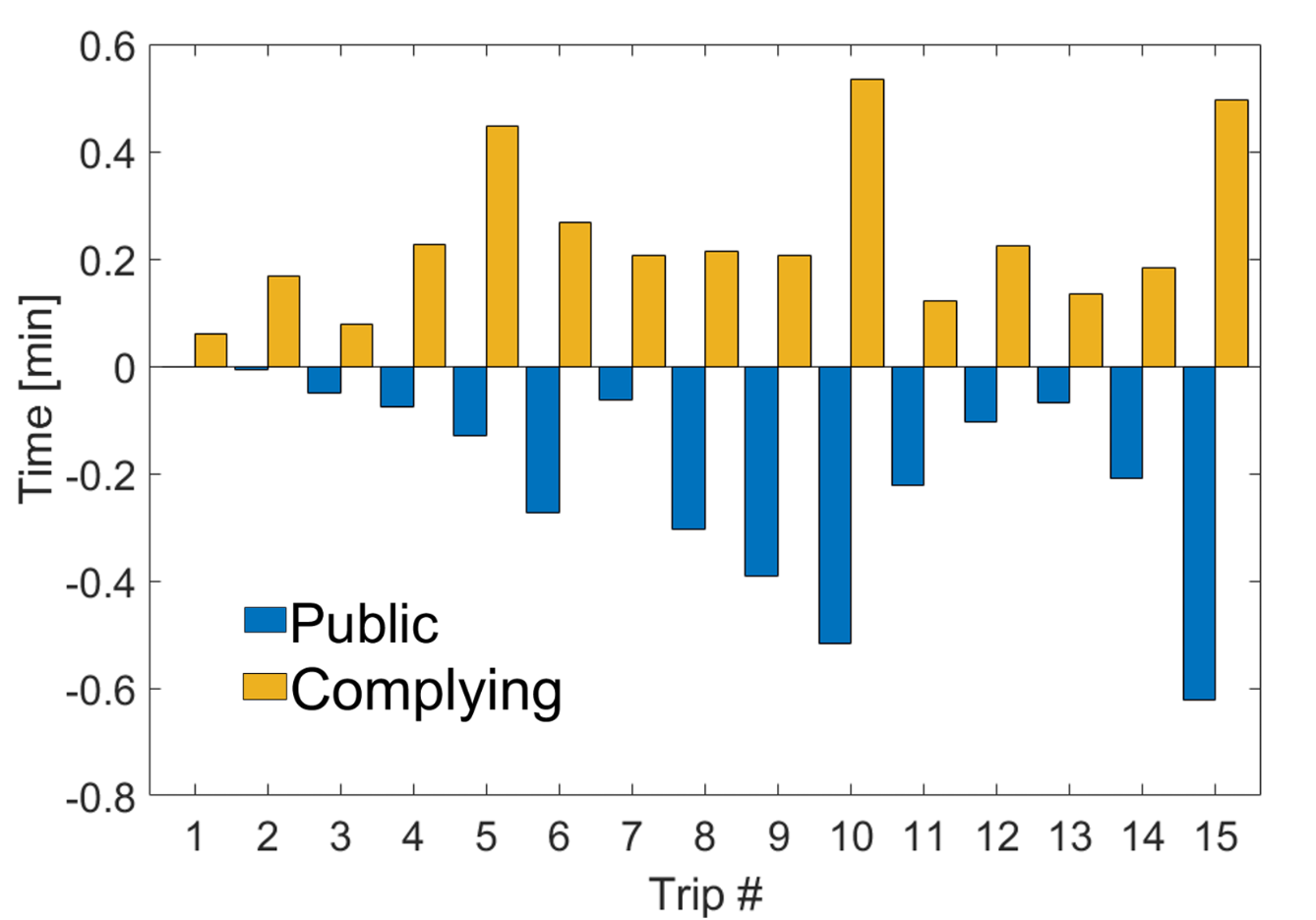}
             \caption{$30\%$ of non-compliance rate}
    \end{subfigure}
    \hfill
    \begin{subfigure}[b]{0.31\textwidth}
        \includegraphics[width=\textwidth]{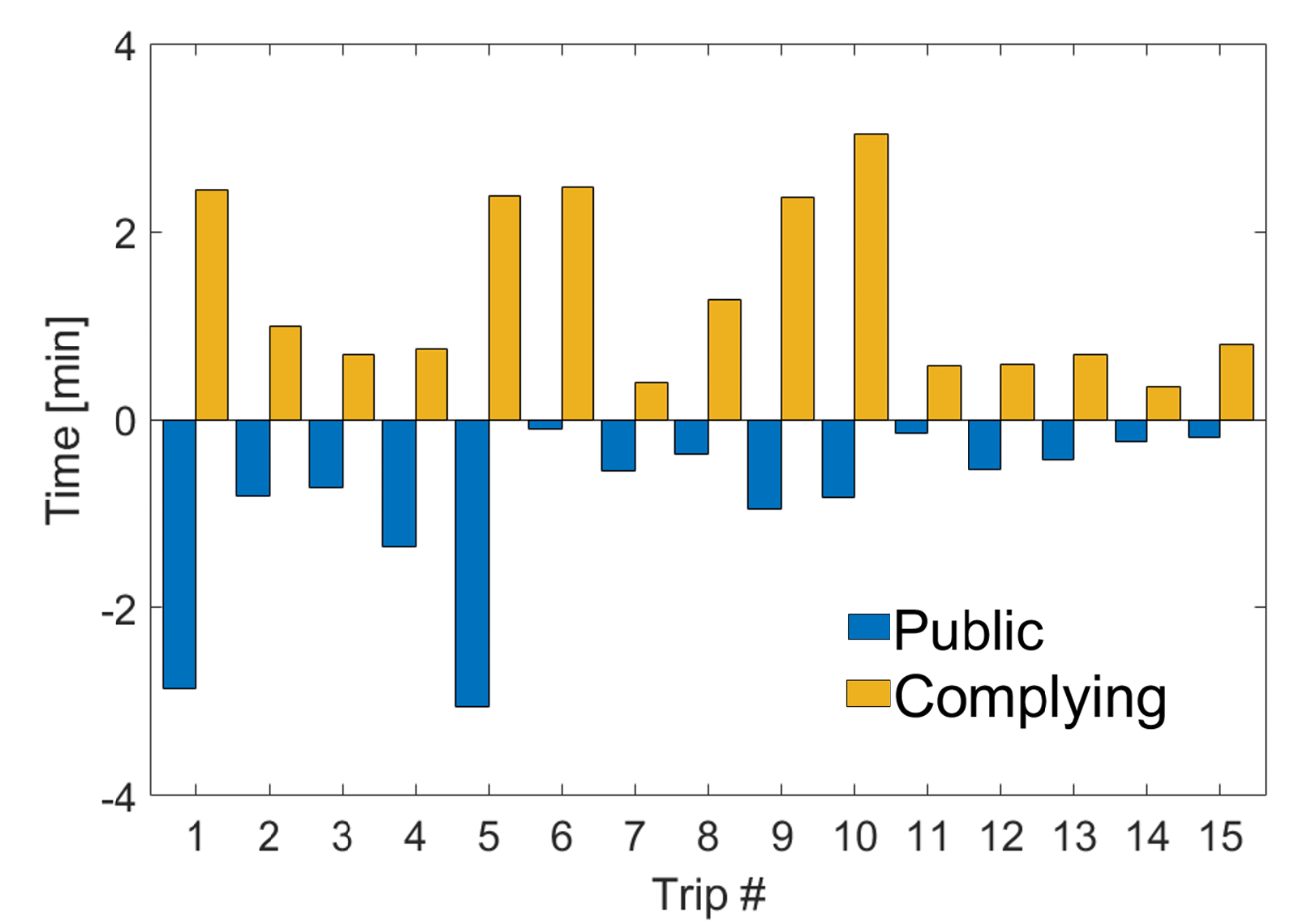}
             \caption{$50\%$ of non-compliance rate}
    \end{subfigure}    

    \caption{Travel-time difference between the first and the second routing iterations for different non-compliance rates with $50\%$ of public transportation and a priority for public transportation of $70\%$.}
    \label{fig:time_difference}
\end{figure*}

Next, we introduce two routing problems: the first seeks to find the system-wide optimal solution for compliant vehicles, and the second models the routing approach followed by non-compliant vehicles with bounded rationality.

\textit{1) System-centric routing} aims to minimize the travel time for all compliant vehicles by solving the following optimization problem for the system planner. 
\begin{problem}[System-Centric Routing] \label{pb:system-centric}
\vspace{3pt}
    \begin{equation}
        \begin{aligned}
            &\minimize_{\{x^{ij}_{m,n}\}} ~ \sum_{m\in\mathcal{M}} w_m \left\{ \sum_{n\in\mathcal{N}}\sum_{(i,j)\in\mathcal{E}} t^{ij}(x^{ij}+q^{ij})\cdot x^{ij}_{m,n} \right\}\\
            \mathrm{s.t.} &~ \sum_{k:(j,k)\in\mathcal{E}} x^{jk}_{m,n} = \alpha_{m,n},~~\forall m\in\mathcal{M},n\in\mathcal{N},j=o_n,\\
            & \sum_{i:(i,j)\in\mathcal{E}} x^{ij}_{m,n} = \alpha_{m,n},~~\forall m\in\mathcal{M},n\in\mathcal{N},j=d_n,\\
            & \sum_{i:(i,j)\in\mathcal{E}} x^{ij}_{m,n} = \sum_{k:(j,k)\in\mathcal{E}} x^{jk}_{m,n},~~\forall m\in\mathcal{M},n\in\mathcal{N},j\in\bar{\mathcal{V}},
        \end{aligned}
    \end{equation}
    where $w_m$ is a weight on mode $m$, $\bar{\mathcal{V}} = \mathcal{V}\setminus\{o_n,d_n\}$ for all $o_n\in\mathcal{O}$, $d_n\in\mathcal{D}$.
\end{problem}

The constraints in Problem \ref{pb:system-centric} ensure that feasible flows are consistent with the demand rate and connect the correct origins to corresponding destinations. Note that Problem \ref{pb:system-centric} is a convex problem as the BPR function is convex in its domain, and the constraints are linear.

\begin{remark}
    The solution to Problem \ref{pb:system-centric} is inherently dependent on the choice of weights $w_m$ for each mode $m \in \mathcal{M}$. Thus, this constitutes a ``low-level" routing problem given the weights.
    The overarching goal of our system planner is to improve MEM. To this end, we present later a problem of selecting optimal weights $w_m$ for each mode $m \in \mathcal{M}$ with this objective.
\end{remark}

\textit{2) Non-compliant vehicle routing} captures the bounded rationality in the decision-making of non-compliant private vehicles using a cognitive hierarchy model \citep{Bang2023mem}.
This model introduces different levels of cognition among human drivers. A driver with a higher cognition level can anticipate the decisions of other drivers with lower cognition levels.
Experimental results imply that humans typically only reason up to level-2 \citep{costa2009comparing,costa2006cognition}; thus, we restrict our model to level-2 decision-making. The cognitive levels are denoted by $\ell = 0,1,2$.

For each trip $n \in \mathcal{N}$, $q_{\ell,n}$ denotes the demand rate for $\ell$-level non-compliant vehicles from the origin $o_n \in \mathcal{O}$ to the corresponding destination $d_n \in \mathcal{D}$.
For an $\ell$-level non-compliant vehicle traveling for trip $n$, we define assignment vector $A_{\ell,n}\in 2^{|\mathcal{E}|}$ where the element $a^{ij}_{\ell,n}$ takes value $1$ if the $\ell$-level non-compliant vehicle for trip $n$ uses the edge $(i,j)$ and takes value $0$ otherwise. Each non-compliant vehicle solves the following problem to determine its path.

\begin{problem}[Non-compliant Vehicle Routing] \label{pb:selfish}
\vspace{3pt}
Each driver with cognition level $\ell\in\{0,1,2\}$ selects the shortest-time path anticipating the flow of lower-level drivers, i.e.,
\begin{equation}
        \begin{aligned}
        \minimize_{\{a_{\ell,n}^{ij}\}}~& \sum_{n\in\mathcal{N}}\sum_{(i,j)\in\mathcal{E}} t^{ij}\left(x^{ij}+\sum_{l=0}^{\ell-1}q_{l}^{ij}\right)\cdot a_{\ell,n}^{ij} \\
        \mathrm{s.t.~} & \sum_{k:(j,k)\in\mathcal{E}} a^{jk}_{\ell,n} = 1,~\forall n\in\mathcal{N},j=o_n,\\
        & \sum_{i:(i,j)\in\mathcal{E}} a^{ij}_{\ell,n} = 1,~\forall n\in\mathcal{N},j=d_n,\\
        & \sum_{i:(i,j)\in\mathcal{E}} a^{ij}_{\ell,n} = \sum_{j:(j,k)\in\mathcal{E}} a^{jk}_{\ell,n},\\
        &\hspace{13ex}\forall n\in\mathcal{N},j\in\mathcal{V}\setminus\{o_n,d_n\},
        \end{aligned}
    \end{equation}
    where $q_\ell^{ij} = \sum_{n\in\mathcal{N}}q_{\ell,n} \cdot a_{\ell,n}^{ij}$ and $q_{\ell,n}$ represents the flow of $\ell$-level non-compliant vehicles for trip $n$.
\end{problem}
Note that $q^{ij}$ is the total flow of non-compliant vehicles on edge $(i,j)$ and $x^{ij}$ in Problems \ref{pb:system-centric} and \ref{pb:selfish} is the flow induced by all the compliant vehicles. We consider that all vehicles can anticipate the flows induced by compliant vehicles through knowledge of public schedules, typical traffic situations at certain hours, etc.

Problem \ref{pb:selfish} is a binary optimization, which we solve using the Dijkstra algorithm.
The solution to Problem \ref{pb:selfish} is the shortest-time path for a single non-compliant vehicle given the compliant vehicles' flow and the lower-level vehicles. Then, we consider all the non-compliant vehicles at the same cognition level for a given origin-destination pair to take the same route because they don't anticipate other vehicles at the same level or their impact on the network.

Solutions to Problem \ref{pb:system-centric} and Problem \ref{pb:selfish} constitute the net flow on the network. This allows us to estimate the travel time $t_m^{o,d}$ for each trip $(o_n,d_n)$, for $n\in\mathcal{N}$, and each mode $m\in\mathcal{M}$.
Consequently, we can evaluate MI at each origin $o_n$ by counting the number of accessible services using mode $m$ within the threshold $\tau_m$, i.e., $\sigma_{o_n,m}^s(\tau_m) = \sum_{o\in\mathcal{O}}\sum_{d\in\mathcal{D}}\mathbb{I}\Big[t_m^{o,d} \leq \tau_m\Big]$, where $\mathbb{I}$ is an indicator function.
Note that the MI resulted from solutions to Problems \ref{pb:system-centric} and \ref{pb:selfish} depends on the weight $w_m$, $m\in\mathcal{M}$.
Thus, we can formulate an optimization problem to obtain better MEM with respect to the weights.


\begin{problem}[Mobility Equity Maximization] \label{pb:mem}
\vspace{3pt}
\begin{equation}
    \begin{aligned}
        \maximize_{w}&~~ \textit{MEM}\\
        \mathrm{subject~to:}&~~ \delta^\mathrm{pv}(w) \leq \gamma,
    \end{aligned}
\end{equation}
where $\delta^\textrm{pv}$ is the average travel-time difference between compliant private vehicles and non-compliant vehicles, and $\gamma$ is the upper limit of the difference.
\end{problem}

We impose the upper limit on the travel-time difference to ensure that the suggested route does not force compliant vehicles to sacrifice their travel time over the limit.

\vspace{-4pt}

\section{Numerical Simulations}  \label{sec:simulation}

\vspace{-2pt}

In this section, we conduct numerical simulations on a grid network to provide insights into how MEM and traffic flow vary with respect to the different parameters. We consider a road network with $16$ nodes and $84$ edges, and the length of each edge is approximately 400 meters. We consider $15$ trips starting from 3 origins to 5 destinations.

As the solutions to Problems \ref{pb:system-centric} and \ref{pb:selfish} affect each other, we solve the problem twice so that they can interact based on the flow induced by other (compliant or non-compliant) vehicles.
As a result, in Fig. \ref{fig:time_difference}, we observe that the travel time of compliant vehicles has increased in the second iteration, while that of public transportation has decreased. This is because compliant private vehicles tend to yield to public transit vehicles.
In addition, the time difference between the first and second iterations increases as the non-compliance rate increases because non-compliant vehicles also benefit from compliant vehicles' yielding behavior. This illustrates the inherent trade-off between the MEM and the travel time gap in Problem \ref{pb:mem}.

\begin{figure*}[h]
    \centering
    \begin{subfigure}[b]{0.28\textwidth}
        \includegraphics[width=\textwidth]{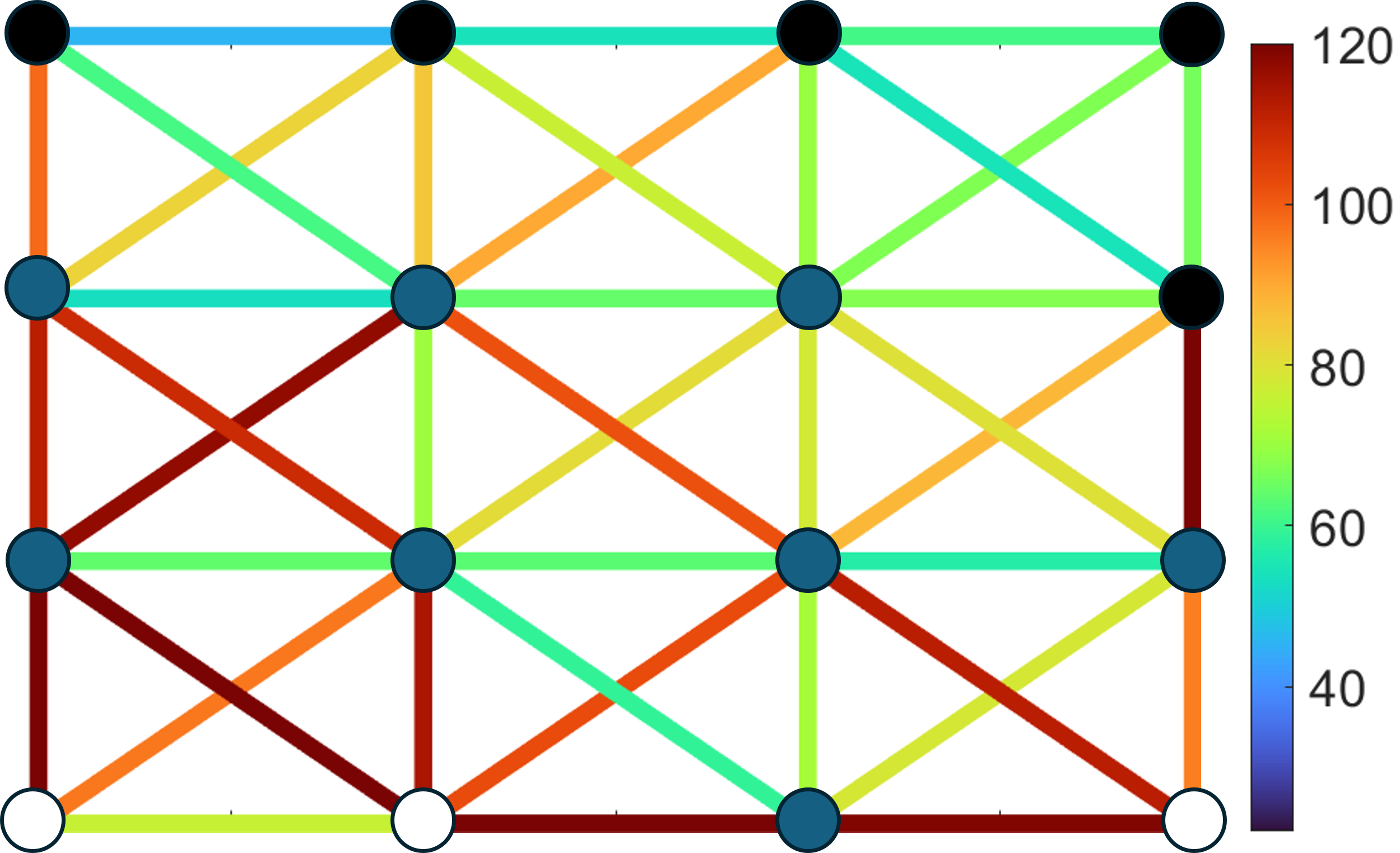}
             \caption{30$\%$ of public transportation}
    \end{subfigure}
    \begin{subfigure}[b]{0.28\textwidth} \hfill
        \includegraphics[width=\textwidth]{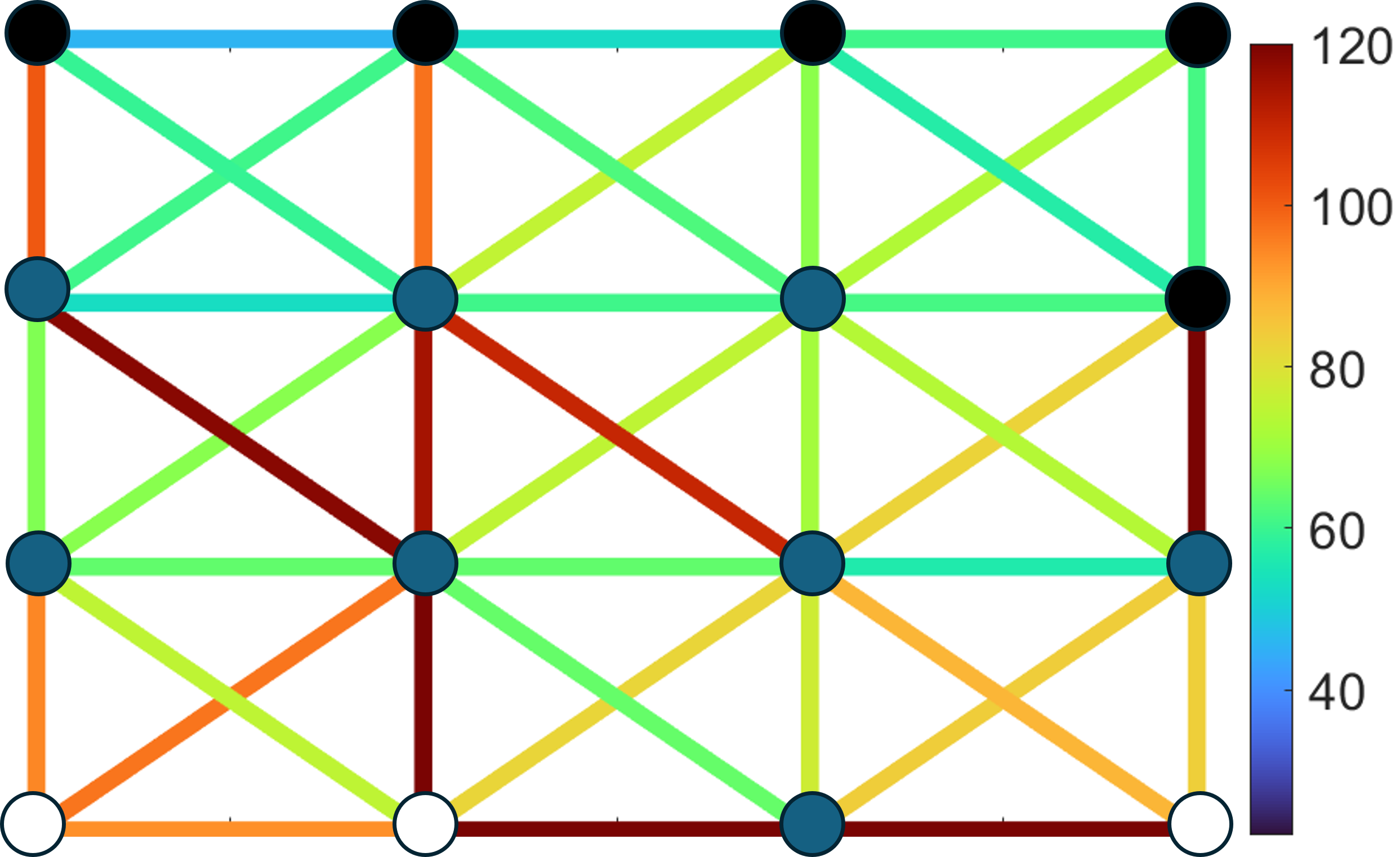}
             \caption{50$\%$ of public transportation}
    \end{subfigure}
    \begin{subfigure}[b]{0.28\textwidth} \hfill
        \includegraphics[width=\textwidth]{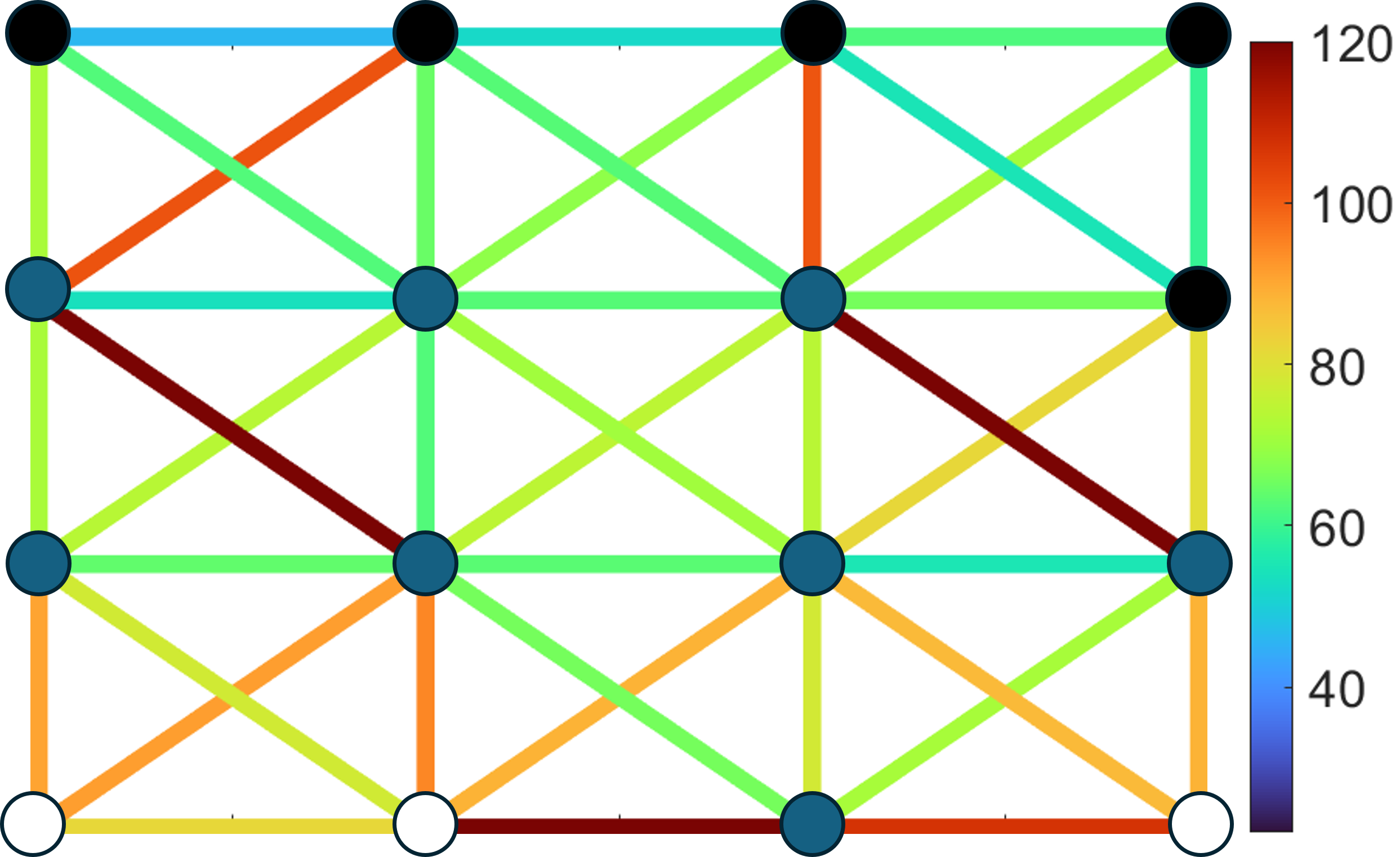}
             \caption{70$\%$ of public transportation}
    \end{subfigure}
    \caption{Travel time per edge for different ratios of public transportation.}
    \label{fig:graph_public_rate}    
\end{figure*}

\begin{figure*}[h]
    \centering
    \begin{subfigure}[b]{0.28\textwidth}
        \includegraphics[width=\textwidth]{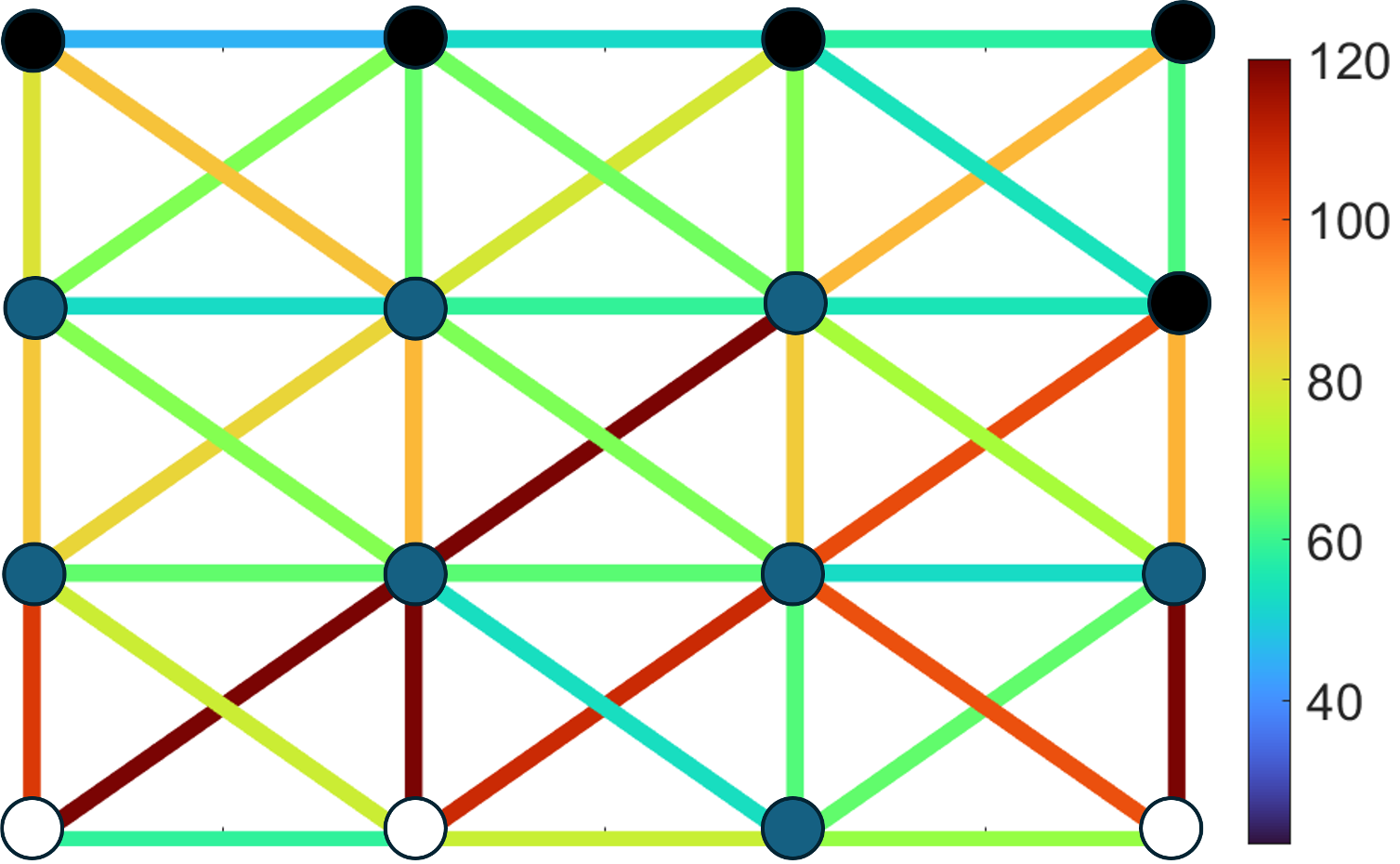}
             \caption{50$\%$ non compliance rate}
    \end{subfigure}
    \begin{subfigure}[b]{0.28\textwidth} \hfill
        \includegraphics[width=\textwidth]{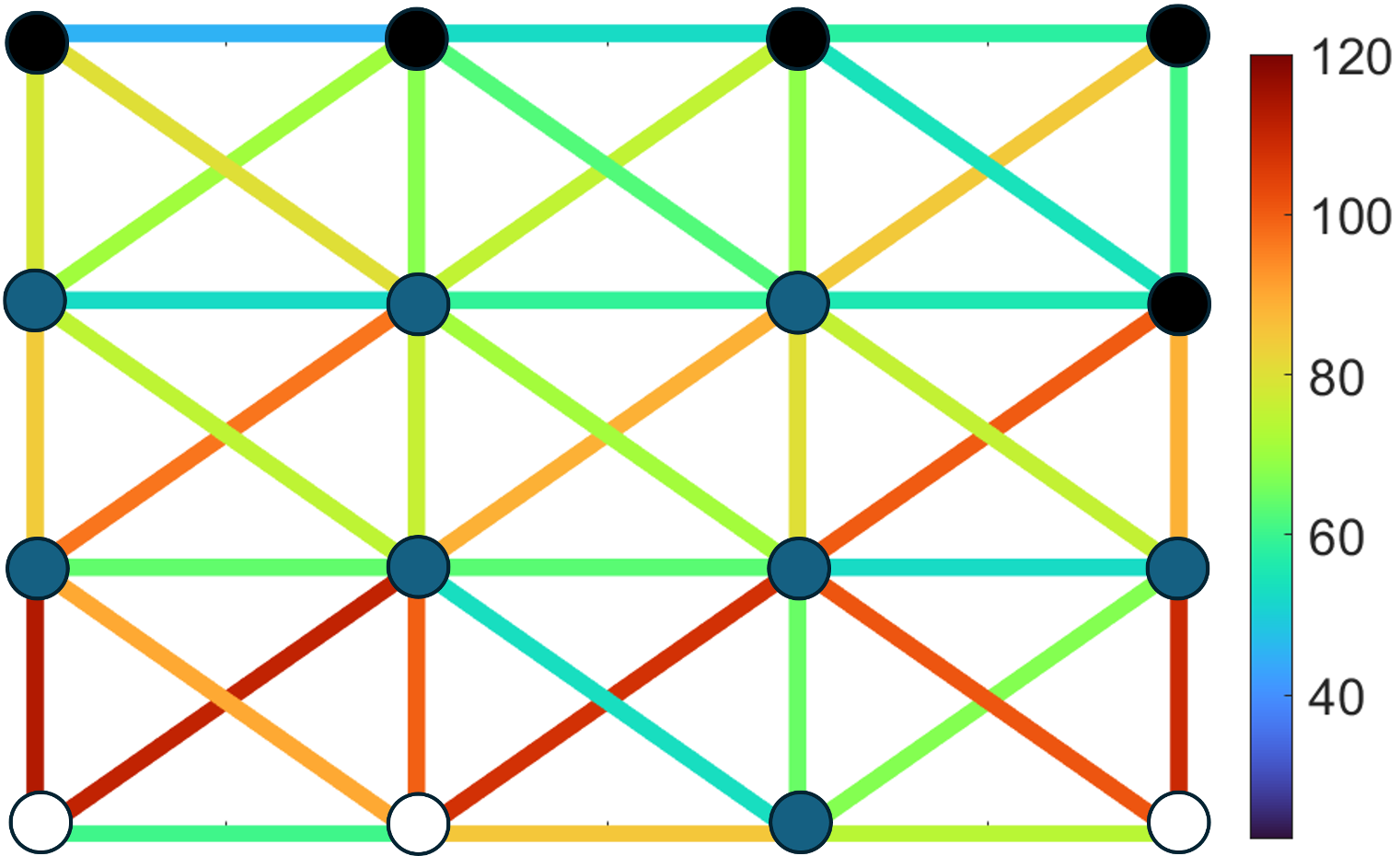}
             \caption{30$\%$ non compliance rate}
    \end{subfigure}
    \begin{subfigure}[b]{0.28\textwidth} \hfill
        \includegraphics[width=\textwidth]{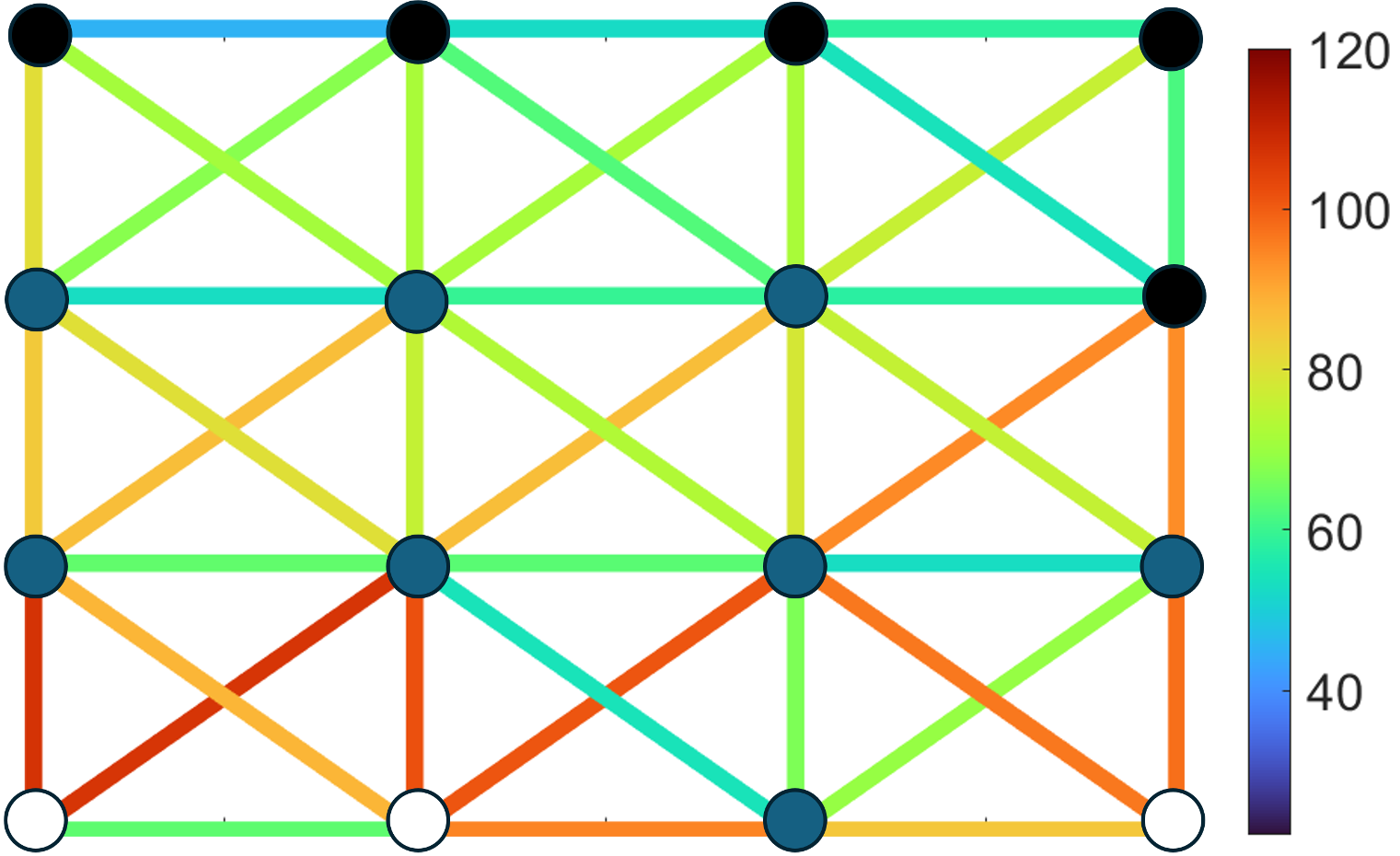}
             \caption{10$\%$ non compliance rate }
    \end{subfigure}
    \caption{Travel time per edge for different ratios of non-compliance rate.}
    \label{fig:graph_compliance}    
\end{figure*}

Next, we present the traffic conditions that result from our solution for different levels of non-compliance and demands for public transit in Figs. \ref{fig:graph_public_rate} and \ref{fig:graph_compliance}. The three white circles are the origins, and the five black circles are the destinations.
The color of the edges represents the travel time across the edge.
In Fig. \ref{fig:graph_public_rate}, we observe that overall travel time decreases with higher usage of public transportation. This is because the net traffic decreases as public transportation tends to accommodate more travelers per unit of traffic than private vehicles.
Likewise, in Fig. \ref{fig:graph_compliance}, as the non-compliance rate decreases, the system planner is able to control more vehicles in the network and, thus, can alleviate congestion on certain edges to improve the travel time throughout the network.

\begin{figure}[h]
    \centering
    \hspace{2pt}\includegraphics[width=0.88\linewidth]{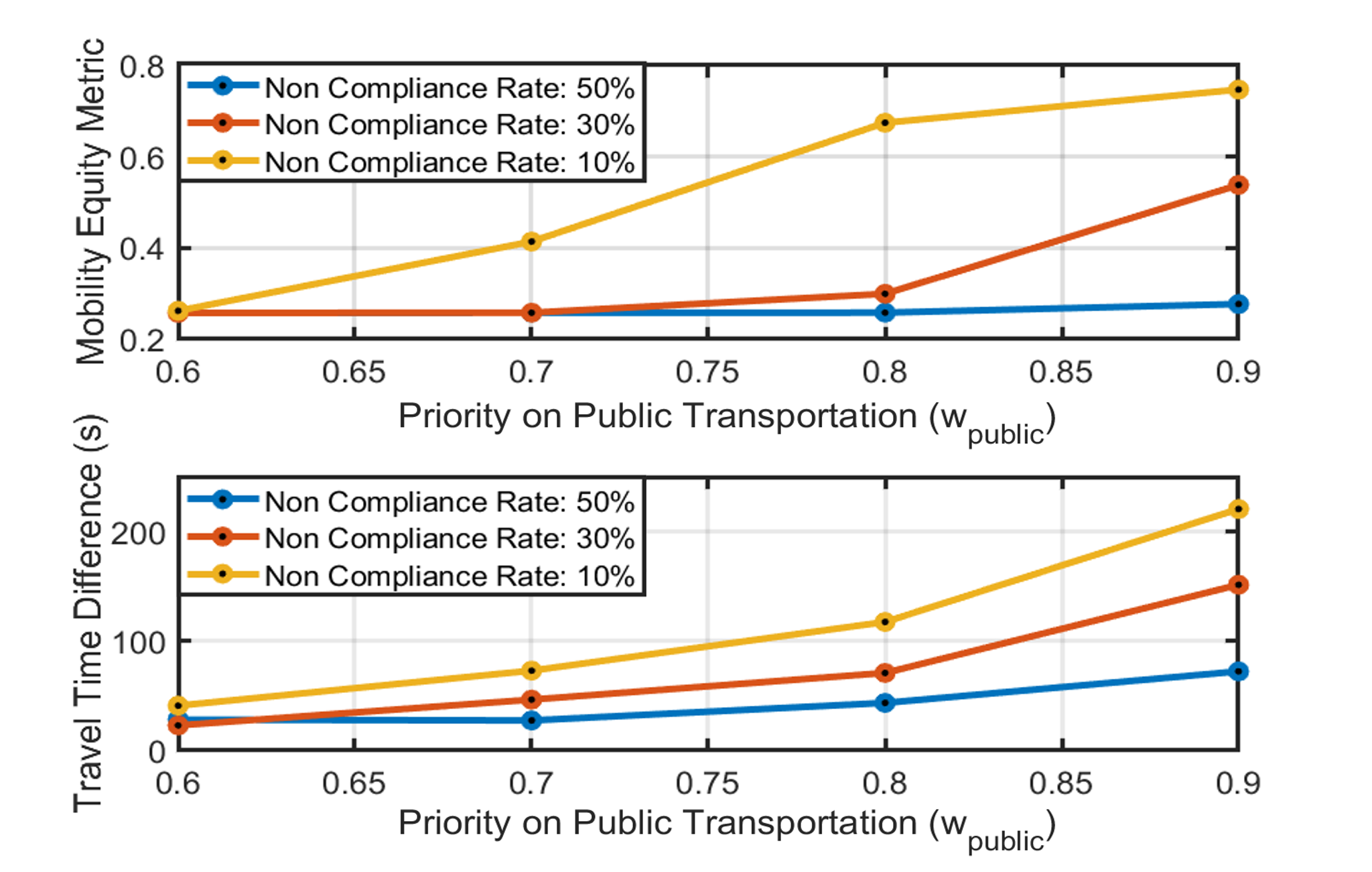}
    \caption{Mobility equity metric and travel time difference with $70\%$ of public transportation.}
    \label{fig:result_ratio_7}
\end{figure}

\begin{figure}[h]
    \centering
    \includegraphics[width=0.88\linewidth]{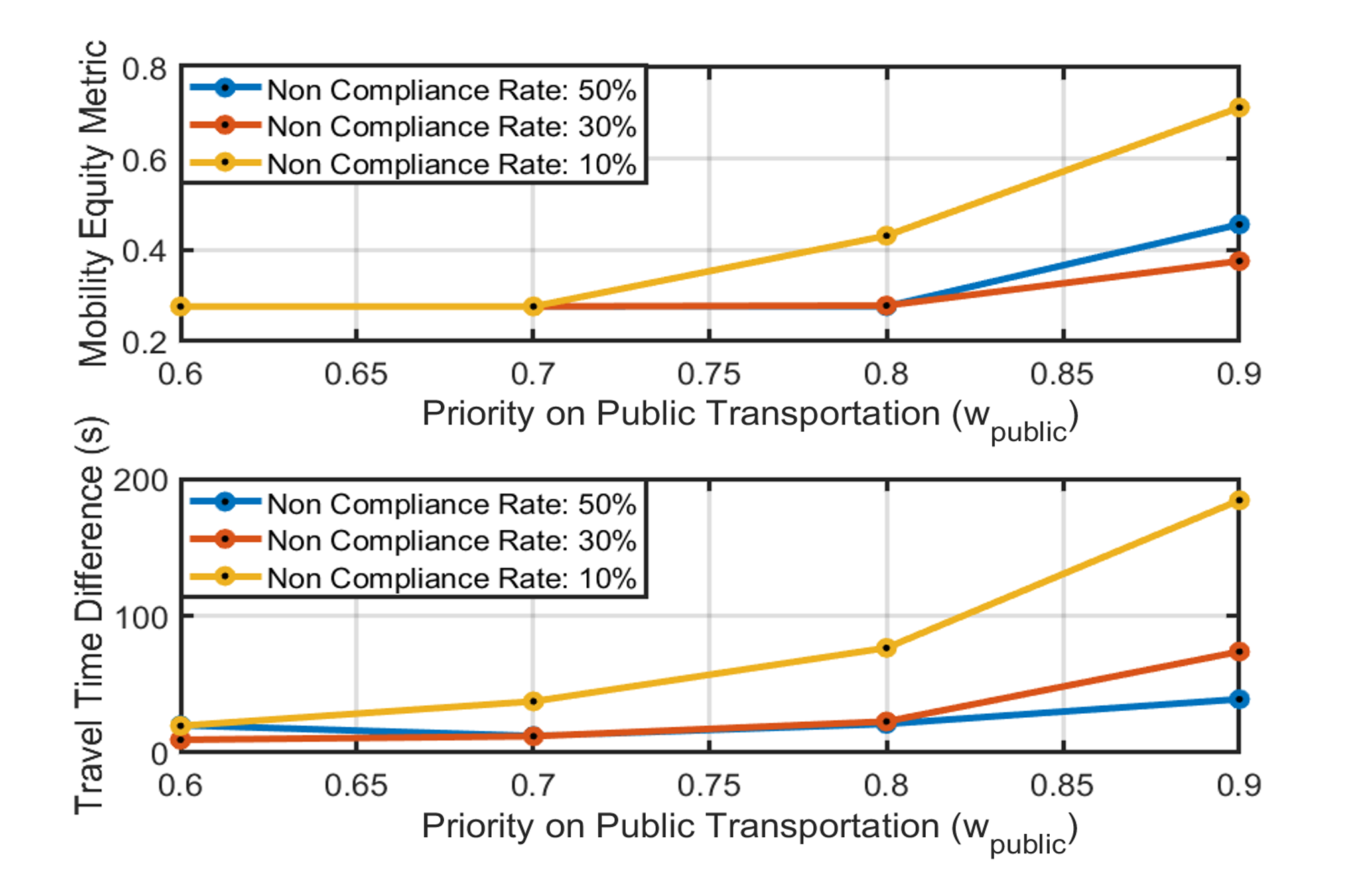}
    \caption{Mobility equity metric and travel time difference with $50\%$ of public transportation.}
    \label{fig:result_ratio_5}
\end{figure}

Lastly, Figs. \ref{fig:result_ratio_7} and \ref{fig:result_ratio_5} visualize how the MEM and travel time difference (between compliant/non-compliant vehicles) vary with respect to the weight assigned to public transportation in Problem \ref{pb:system-centric}. For example, MEM and travel time difference increase along with the weight on public transportation, and in practice, this result provides insights on designing proper constraints in Problem \ref{pb:mem}.

\section{Discussion and Conclusions} \label{sec:conclusion}

In this paper, we introduced a MEM that measures the equity in the distribution of the ability to move and
provided an optimization framework that maximizes MEM in transportation networks.
Through numerical simulations, we analyzed the impact of MEM optimization on public transit, as well as compliant and non-compliant vehicles.
Future work should analyze the implications of MEM in a real city network.

\bibliography{Bang,IDS,MEM}

\begin{thebibliography}{19}
\providecommand{\natexlab}[1]{#1}
\providecommand{\url}[1]{\texttt{#1}}
\providecommand{\urlprefix}{URL }
\expandafter\ifx\csname urlstyle\endcsname\relax
  \providecommand{\doi}[1]{doi:\discretionary{}{}{}#1}\else
  \providecommand{\doi}{doi:\discretionary{}{}{}\begingroup \urlstyle{rm}\Url}\fi

\bibitem[{Ammous et~al.(2017)Ammous, Belakaria, Sorour, and Abdel-Rahim}]{ammous2017optimal}
Ammous, M., Belakaria, S., Sorour, S., and Abdel-Rahim, A. (2017).
\newblock Optimal routing with in-route charging of mobility-on-demand electric vehicles.
\newblock In \emph{2017 IEEE 86th Vehicular Technology Conference (VTC-Fall)}, 1--5. IEEE.

\bibitem[{Au et~al.(2015)Au, Zhang, and Stone}]{Au2015}
Au, T.C., Zhang, S., and Stone, P. (2015).
\newblock {Autonomous intersection management for semi-autonomous vehicles}.
\newblock \emph{Handbook of Transportation, Routledge, Taylor \& Francis Group,}.

\bibitem[{Bang et~al.(2023)Bang, Dave, and Malikopoulos}]{Bang2023mem}
Bang, H., Dave, A., and Malikopoulos, A.A. (2023).
\newblock {Routing in Mixed Transportation Systems for Mobility Equity}.
\newblock \emph{Proceedings of the 2023 American Control Conference, (to appear, arXiv:2309.03981)}.

\bibitem[{Bang et~al.(2024)Bang, Dave, and Malikopoulos}]{bang2024confidence}
Bang, H., Dave, A., and Malikopoulos, A.A. (2024).
\newblock Safe merging in mixed traffic with confidence.
\newblock \emph{(in review, arXiv:2403.05742)}.

\bibitem[{Bang and Malikopoulos(2022)}]{bang2021AEMoD}
Bang, H. and Malikopoulos, A.A. (2022).
\newblock Congestion-aware routing, rebalancing, and charging scheduling for electric autonomous mobility-on-demand system.
\newblock In \emph{Proceedings of 2022 American Control Conference (ACC)}, 3152--3157.

\bibitem[{Costa-Gomes and Crawford(2006)}]{costa2006cognition}
Costa-Gomes, M.A. and Crawford, V.P. (2006).
\newblock Cognition and behavior in two-person guessing games: An experimental study.
\newblock \emph{American economic review}, 96(5), 1737--1768.

\bibitem[{Costa-Gomes et~al.(2009)Costa-Gomes, Crawford, and Iriberri}]{costa2009comparing}
Costa-Gomes, M.A., Crawford, V.P., and Iriberri, N. (2009).
\newblock Comparing models of strategic thinking in van huyck, battalio, and beil's coordination games.
\newblock \emph{Journal of the European Economic Association}, 7(2-3), 365--376.

\bibitem[{Deboosere and El-Geneidy(2018)}]{deboosere2018evaluating}
Deboosere, R. and El-Geneidy, A. (2018).
\newblock Evaluating equity and accessibility to jobs by public transport across canada.
\newblock \emph{Journal of Transport Geography}, 73, 54--63.

\bibitem[{Gastwirth(1972)}]{gastwirth1972estimation}
Gastwirth, J.L. (1972).
\newblock The estimation of the lorenz curve and gini index.
\newblock \emph{The review of economics and statistics}, 306--316.

\bibitem[{Guo et~al.(2020)Guo, Chen, Stuart, Li, and Zhang}]{guo2020systematic}
Guo, Y., Chen, Z., Stuart, A., Li, X., and Zhang, Y. (2020).
\newblock A systematic overview of transportation equity in terms of accessibility, traffic emissions, and safety outcomes: From conventional to emerging technologies.
\newblock \emph{Transportation research interdisciplinary perspectives}, 4, 100091.

\bibitem[{Li et~al.(2018)Li, Kolmanovsky, Girard, and Yildiz}]{li2018game}
Li, N., Kolmanovsky, I., Girard, A., and Yildiz, Y. (2018).
\newblock Game theoretic modeling of vehicle interactions at unsignalized intersections and application to autonomous vehicle control.
\newblock In \emph{2018 Annual American Control Conference (ACC)}, 3215--3220. IEEE.

\bibitem[{Litman(2017)}]{litman2017evaluating}
Litman, T. (2017).
\newblock \emph{Evaluating transportation equity}.
\newblock Victoria Transport Policy Institute Victoria, BC, Canada.

\bibitem[{Salazar et~al.(2019)Salazar, Tsao, Aguiar, Schiffer, and Pavone}]{salazar2019congestion}
Salazar, M., Tsao, M., Aguiar, I., Schiffer, M., and Pavone, M. (2019).
\newblock A congestion-aware routing scheme for autonomous mobility-on-demand systems.
\newblock In \emph{2019 18th European Control Conference (ECC)}, 3040--3046. IEEE.

\bibitem[{Ta et~al.(2017)Ta, Li, Zhao, Feng, Ma, and Gong}]{ta2017efficient}
Ta, N., Li, G., Zhao, T., Feng, J., Ma, H., and Gong, Z. (2017).
\newblock An efficient ride-sharing framework for maximizing shared route.
\newblock \emph{IEEE Transactions on Knowledge and Data Engineering}, 30(2), 219--233.

\bibitem[{TOMTOM(2023)}]{TOMTOM}
TOMTOM (2023).
\newblock Pointes of interest search.
\newblock \urlprefix\url{https://developer.tomtom.com/search-api/ documentation/search-service/points- of-interest-search}.
\newblock Accessed: 2023-04-01.

\bibitem[{Tsao et~al.(2019)Tsao, Milojevic, Ruch, Salazar, Frazzoli, and Pavone}]{tsao2019model}
Tsao, M., Milojevic, D., Ruch, C., Salazar, M., Frazzoli, E., and Pavone, M. (2019).
\newblock Model predictive control of ride-sharing autonomous mobility-on-demand systems.
\newblock In \emph{2019 International Conference on Robotics and Automation (ICRA)}, 6665--6671. IEEE.

\bibitem[{Wollenstein-Betech et~al.(2022)Wollenstein-Betech, Salazar, Houshmand, Pavone, Paschalidis, and Cassandras}]{wollenstein2021routing}
Wollenstein-Betech, S., Salazar, M., Houshmand, A., Pavone, M., Paschalidis, I.C., and Cassandras, C.G. (2022).
\newblock Routing and rebalancing intermodal autonomous mobility-on-demand systems in mixed traffic.
\newblock \emph{IEEE Transactions on Intelligent Transportation Systems}, 23(8).

\bibitem[{Yu et~al.(2021)Yu, Miao, Bayram, Yu, and Chen}]{yu2021optimal}
Yu, X., Miao, H., Bayram, A., Yu, M., and Chen, X. (2021).
\newblock Optimal routing of multimodal mobility systems with ride-sharing.
\newblock \emph{International Transactions in Operational Research}, 28(3), 1164--1189.

\bibitem[{Zhou et~al.(2019)Zhou, Yu, and Qu}]{zhou2019development}
Zhou, M., Yu, Y., and Qu, X. (2019).
\newblock Development of an efficient driving strategy for connected and automated vehicles at signalized intersections: A reinforcement learning approach.
\newblock \emph{IEEE Transactions on Intelligent Transportation Systems}, 21(1), 433--443.

\end{thebibliography}

\end{document}